# The electronic structure of poly(pyridine-2,5-diyl) investigated by soft x-ray absorption and emission spectroscopies

M. Magnuson[1], L. Yang[2], J.- H. Guo[1], C. Såthe[1], A. Agui[1], J. Nordgren[1],
Y. Luo[3], H. Ågren[2], N. Johansson[2], W. R. Salaneck[2], L. E. Horsburgh[4]
and A. P. Monkman[4]

[1]*Department of Physics, Uppsala University, Box 530, S-75121 Uppsala, Sweden*

[2]*Institute of Physics and Measurement Technology, Linköping University, S-58183, Linköping, Sweden*

[3]*FYSIKUM, University of Stockholm, Box 6730, S-113 85 Stockholm, Sweden*

[4]*Department of Physics, University of Durham, South Road, Durham DH1 2UE, England*

**Abstract**

The electronic structure of the poly-pyridine conjugated polymer has been investigated by resonant and nonresonant inelastic X-ray scattering and X-ray absorption spectroscopies using synchrotron radiation. The measurements were made for both the carbon and nitrogen contents of the polymer. The analysis of the spectra has been carried out in comparison with molecular orbital calculations taking the repeat-unit cell as a model molecule of the polymer chain. The simulations indicate no significant differences in the absorption and in the non-resonant X-ray scattering spectra for the different isomeric geometries, while some isomeric dependence of the resonant spectra is predicted. The resonant emission spectra show depletion of the π electron bands in line with symmetry selection and momentum conservation rules. The effect is most vizual for the carbon spectra; the nitrogen spectra are dominated by lone pair *n* orbital emission of σ symmetry and are less frequency dependent.

## 1 Introduction

Conjugated polymers have been the subject of much interest owing to their unique electronic properties which can be technically exploited, as, for example, for doping induced electrical conductors and light emitting diodes [1]. Detailed experimental studies of the uppermost π-valence levels at the valence and conducting band edges of such polymers are important for the understanding of these properties. Such studies have been carried out by many techniques, for instance by photoelectron spectroscopy using photon excitation in both the X-ray and ultraviolet wavelength regimes [2-4].

The purpose of the present work is to use X-ray emission (XE) spectroscopy to investigate the electronic structure of poly(pyridine-2,5-diyl) (PPy) which is an aza-substituted poly(*p*-phenylene) with a sufficiently large bandgap to allow electroluminescence in the blue wavelength region [5, 6]. X-ray emission spectroscopy provides an alternative technique for studying polymers but has been quite little exploited so far in that context [15]. It provides a means of extracting the electronic structure information in terms of local contributions to the Bloch or molecular valence orbitals, since the decay from the core-hole can be described according to local dipole selection rules [7,8]. The method is atomic element specific and also symmetry selective at high resolution. However, the relatively low fluorescence yield and instrument efficiencies associated with XE in the ultra-soft x-ray regime makes a considerable demand on the experimental measurements in this spectral region. An intense, *synchrotron radiation* (SR), excitation source is therefore required which has limited the activity in this field.

In a recent work we used some poly-phenylene-vinylenes, PPV, PDPV and PMPV, to demonstrate the possibility to produce resonant and nonresonant X-ray scattering spectra of polymer compounds by means of monochromatic synchrotron radiation [15]. The spectra were sufficiently resolved to allow an analysis of the spectral features in terms of electronic structure theory, and also to analyze the salient changes between the non-resonant to resonant spectra in terms of the differences of the underlying physical processes. The nonresonant as well as the resonant spectra measured in Ref. [15] demonstrated for each polymer benzene-like features, indicating a local character of the X-ray emission in which the phenyl ring acts as a building block. The edges of the occupied bands could be identified in each case allowing for an alternative way of determining the optical band gap of the polymers, namely that the subtraction of the measured core-excitation (LUMO) resonance and the non-resonant HOMO emission energies.

The most conspicuous feature of the polymer spectra reported in Ref. [15] was, however, that, like for free benzene, the emission from the outer $\pi$ band was depleted going from the non-resonant to the LUMO-resonant X-ray emission spectra. It was demonstrated that this transition(s), which is strictly symmetry forbidden for free benzene, becomes effectively forbidden, - and momentum conserving-, in the polymer case as a result of strong interference effects, and it was argued that this is the general case for resonant X-ray emission of conjugated polymers as far as the frozen orbital approximation holds. The poly-phenylene-vinylenes are simple hydrocarbon polymers, and it is of interest to find out how the resonant and non-resonant spectra show up in more complicated hetero-compounds, then also mapping the energy bands by transitions from more than one atomic element. In the present work we present, and analyze for this purpose, X-ray spectra of poly-pyridine which is an aza-substituted poly-phenylene. This also provides the possibility to study the isomeric dependence of the spectra both in emission and absorption.

In the following section we describe the details of the experiment carried out at the Lawrence Berkeley National Laboratory (section II), followed by some aspects of the simulations (section III). In section IV, the results for non-resonant and resonant nitrogen and carbon spectra are presented and analyzed with respect to the simulations. The role of different isomers of poly-pyridine is then also considered. The corresponding X-ray absorption (XA) spectra are also analyzed and included in the simulations. The results are discussed and summarized in the final sections V and VI.

## 2 Experimental Details

The experiment was carried out at beamline 7.0 at the ALS. This undulator beamline includes a spherical-grating monochromator [16] and provides linearly polarized SR of high resolution and high brightness. X-ray absorption (NEXAFS) spectra were recorded by measuring the total electron yield from the sample current with 0.25 eV and 0.40 eV resolution of the beamline monochromator for the carbon and nitrogen edges, respectively. The NEXAFS spectra were normalized to the incident photon current using a clean gold mesh inserted in the excitation beam.

The X-ray emission (XE) spectra were recorded using a high-resolution grazing-incidence x-ray fluorescence spectrometer [17]. During the XE measurements, the resolution of the beamline monochromator was the same as in the NEXAFS measurements. The x-ray fluorescence spectrometer had a resolution of 0.30 eV and 0.65 eV, for the carbon and nitrogen measurements, respectively. The energy scale has been calibrated using the elastic (recombination) peak in the XE spectra which has the same energy as the incoming photon energy. The sample was oriented so that the incidence angle of the photons was 20 degrees with respect to the surface plane. During the data collection, the samples were scanned (moved every 30 seconds) with respect to the photon beam to avoid the effects from photon-induced decomposition of the polymer. The base pressure in the experimental chamber was $4 \times 10^{-9}$ Torr during the measurements.

The polymer was synthesised following a modified Yamamoto route [18] and spin-coated from formic acid solution to yield thick films on optically flat Si(110) wafers. The samples were sealed in an $N_2$ atmosphere in glass tubes and exposed to air for a very limited time just before they were introduced into a vacuum system and mildly baked at $150^o$ C for about 12 hours, to minimize water content. PPy is however stable in air and does not readily oxidize, therefore this treatment will not cause any ill effects.

## 3 Calculations

The X-ray absorption and inelastic X-ray scattering (or inelastic X-ray emission) processes can be expressed by the following relations;

$$A + \hbar\omega \rightarrow A(k^{-1}\nu) \quad (P1)$$

$$\rightarrow A(n^{-1}\nu) + \hbar\omega' \quad (P2)$$

where $k$, $n$ and $\nu$ denote levels defined by the core, occupied valence, and unoccupied molecular orbitals (MO's). The first step (P1), refers to the X-ray absorption process with an electron being excited from the core orbital $k$ to a virtual MO $\nu$, where $\nu$ can be both a discrete or a continuum level. When the core hole is filled by an electron from the occupied valence MO with an emission of a photon as shown by process P2, the inelastic X-ray scattering process occurs. Furthermore, if $\nu$ is a discrete MO (for example $\pi^*$) in process P2, a resonant inelastic X-ray scattering (RIXS) spectrum is observed. If $\nu$ denotes a continuum level, a non-resonant inelastic X-ray scattering spectrum is obtained instead.

The analysis of the RIXS process requires in general a one-step formalism, which leads to a Kramers-Heisenberg-type dispersion formula for the cross section. The general theory for a RIXS spectrum of randomly oriented molecular and polymeric systems have been presented in previous works[15,9,19,20,21] and is also applied for this study. For the general description of calculations of non-resonant inelastic X-ray scattering and X-ray absorption spectra (NEXAFS), we also refer to earlier studies [15,9,10,12,13], and to Ref. [14] for a more rigorous *ab initio* band theory formulation of the non-resonant X-ray emission spectra, the implementation of which, however, is limited by the applicable size of the unit cell.

Figure 1 gives an illustration of the head-to-head (HH) and head-to-tail (HT) unit cells of the four different isomers possible in polypyridine. An extra letter, T or C (trans or cis), denotes the relative positions between adjacent nitrogen atoms. The simulations were carried out by taking a repeat-unit (dimer) cell as a model molecule of the poly-pyridine polymer which includes two pyridine rings. The restriction to one single repeat unit finds motivation since poly-pyridine has relatively flat bands. Few test calculations on multiple-unit spectra did not give significant differences. The geometries of the model molecules were obtained by using the AM1 Hamiltonian [22] in the MOPAC program for a bigger model system (a model molecule with 4 pyridine rings). Calculations of the pyridine molecule was also carried out for comparative purpose. The direct self-consistent field (SCF) program DISCO [25], modified for static exchange (STEX) calculations [10], has been employed for the orbital energy and dipole transition moment calculations.

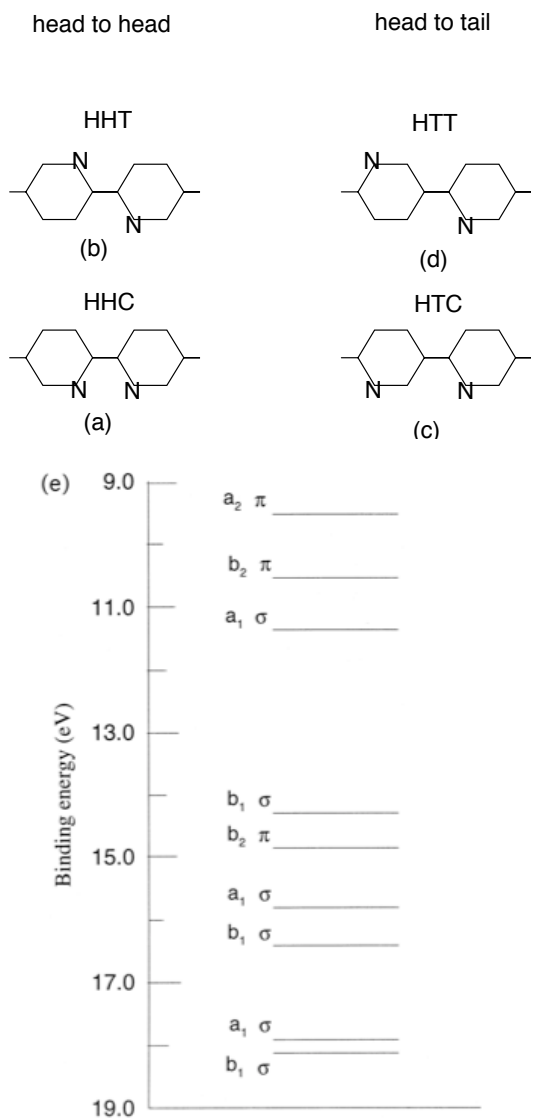

**Fig. 1:** Panels a,b,c,d: The isomeric head-to-tail (H-T) and head-to-head (H-H) geometries of PPy; The extra letter, T and C (*trans* and *cis*) denotes the relative positions between adjacent nitrogen atoms (180 and 0 degrees torsion angle). Panel 1-e: Orbital energy diagram of pyridine.

The STEX ab initio method, extensively described elsewhere [10,11] and adopted in the present calculations, is based on an independent particle description of initial and final states in the photon absorption process. The initial state is then the SCF ground state while the final states are described as configurations due to single excitations from the core orbital to a set of optimized virtual orbitals. These are eigenvectors of a one-particle Hamiltonian that describes the motion of the excited electron in the field of the remaining molecular ion including the electron relaxation around the core-hole. In our approach the STEX Hamiltonian matrix is constructed directly from one- and two-electron integrals computed in the atomic orbital basis set by modifying the electron density. The bound molecular orbitals (MO) from SCF and ΔSCF optimizations are expanded in a relatively small basis set, here a standard polarization basis set [26] (DZP), while the construction of the Hamiltonian matrix is obtained in an augmented, very large, but non-redundant gaussian basis set, centered at the atom with the core hole. (The exponents are generated by the formula $\alpha_n = \alpha_0 * \beta^{-n}$, n=0,1,2,,$n_{max}$, where $n_{max}$ = 14,15,14 for s,p,d, respectively, and $\alpha_0$=0.0928, $\beta$=1.3.) The occupied MOs are then projected out from the augmented basis set, and the projected STEX Hamiltonian is diagonalized completely to yield eigenvalues and eigenvectors. This algorithm has been implemented in a modified version of the program DISCO of Almlöf and co-workers [25]. The eigenpairs of excitation energies and corresponding oscillator strengths provide a primitive spectrum for a Stieltjes imaging procedure in the continuum in order to obtain the averaged ionization cross section. The total NEXAFS spectra have been computed with the assumption that the "discrete" excited states at energies above the first ionization threshold can be considered as autoionizing states and their intensity merges in the continuum of the open channels.

# 4 Results

## 4.1 X-ray absorption spectra

### 4.1.1 $N_{1s}$ X-ray absorption

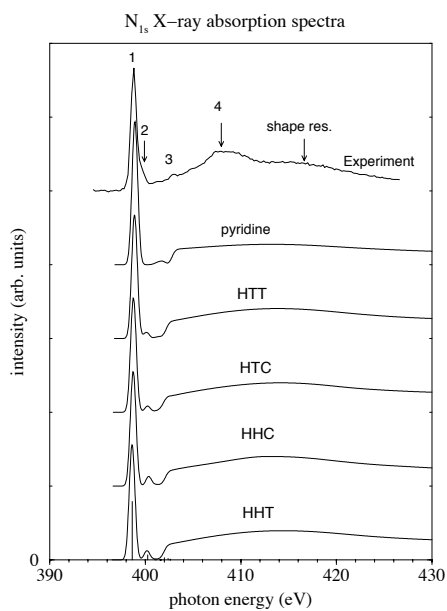

**Fig. 2:** Experimental and calculated XA spectra of PPy at the $N_{1s}$ threshold, where the calculated spectra have been aligned to the $\pi^*$ peak of the experimental spectrum.

Figure 2 shows an experimental X-ray absorption spectrum measured at the nitrogen 1s threshold of polypyridine. An intense peak at about 399 eV corresponds to the lowest unoccupied $\pi^*$ molecular orbital (LUMO). At higher excitation energies, shape resonances mainly due to the $\sigma^*$ levels and multielectron processes are observed. Calculated absorption spectra are also presented for the four poly-pyridine isomers. Table 1 lists the assignments of the absorption features in the discrete energy spectrum.

For a qualitative understanding of the different transitions involved, it is instructive to compare with calculations of the pyridine molecule. The ground state of the pyridine molecule has $C_{2v}$ group symmetry and its two lowest unoccupied MO's are labeled $3b_1$ (LUMO) and $2a_2$ (LUMO+1). Due to symmetry, the $N_{1s}$ core to $2a_2$ MO transition lacks intensity. For a poly-pyridine unit-cell model molecule, the two lowest unoccupied MO's (LUMO and LUMO+1) are both of $\pi$ character which essentially corresponds to the $3b_1$ and $2a_2$ MO's of the pyridine subunits. However, unlike pyridine, the LUMO+1 transition is accessible in poly-pyridine due to the C-C bridge-bonding of the two subunits which lowers the symmetry of the system.

In the calculated spectrum, the $N_{1s} \rightarrow$ LUMO+1 transition appears as a peak in the discrete energy region and can be observed in the experimental $N_{1s}$ spectrum as a high-energy shoulder of the main peak ($\pi^*$). A population quantum defect analysis assigns a quite large 3p character. The third feature in the experimental spectrum falls close to the computed ionization potential. A shape resonance centered at about 11.5 eV higher than the $N_{1s}$ ionization potential can also be observed. These resonances can be attributed to those observed by inner shell electron energy loss spectroscopy (ISEELS) measurements of the $N_{1s}$ excitation of pyridine[28]. The features are shifted to higher energy positions in poly-pyridine compared to the pyridine molecule by about 2.0 eV. The calculated absorption spectra indicate no significant difference for the four poly-pyridine isomers.

**Table 1:** Assignment of the nitrogen and carbon K-edge XA structures in PPy. CC,CN,CH denote a bridge carbon, and ring carbons connected and unconnected to a nitrogen, respectively.

| Feature | Energy (eV) $N_{1s}$ | Assignment | Energy (eV) $C_{1s}$ | Assignment |
|---|---|---|---|---|
| 1 | 398.8 | $\pi^*$ | 284.4 | $\pi^*(CH)$ |
| 1' |  |  | 285.0 | $\pi^*(CN;CC)$ |
| 2 | 399.5 | $3p\pi(2a_2)$ | 287.5 | $(3p+3d)\sigma$ |
| 3 | 403.0 | *IP* | 289.2 | *IP* |
| 4 | 407.5 | *Multielectron* | 293.3 | *Multielectron* |

### 4.1.2 $C_{1s}$ X-ray absorption

Figure 3 shows an experimental X-ray absorption spectrum measured at the carbon 1s threshold of poly-pyridine. A double peak structure at about 285 eV corresponds to the lowest unoccupied $C_{1s}$ (LUMO) orbital which is core-level split due to different chemical environments for the carbon atoms. The figure also shows the calculated absorption spectra at the $C_{1s}$ threshold. The assignments of the absorption features are given in Table 1. The largest difference between the calculated absorption spectra at the $N_{1s}$ and the $C_{1s}$ thresholds (see Fig. 2 and Fig. 3) is the splitting of the most intensive feature in the $C_{1s}$ spectrum (noted as 1 and 1', respectively, in Table 2). Three extra transitions in the $C_{1s}$ absorption spectrum can also be observed. Similar shape resonances as in the nitrogen case are observed in the carbon absorption spectrum. The relative position of the shape resonance is about the same as for the nitrogen spectra. From Fig. 2 and 3, it is clear that the calculated absorption spectra at the $N_{1s}$ and the $C_{1s}$ thresholds do not show any large dependence on the isomeric geometries, thus indicating that the isomeric character of polymers of this kind is hard to probe with X-ray absorption spectroscopy [29]. The double peak structure of the $\pi^*$ band in the carbon spectrum is obviously due to chemical shifts. The corresponding pyridine spectrum also show a double peak, which is assigned to the carbons connected and unconnected to the nitrogen. The same is true also for the polypyridine double-ring model; the simulations favour a head-to-head geometry but are not conclusive on that point.

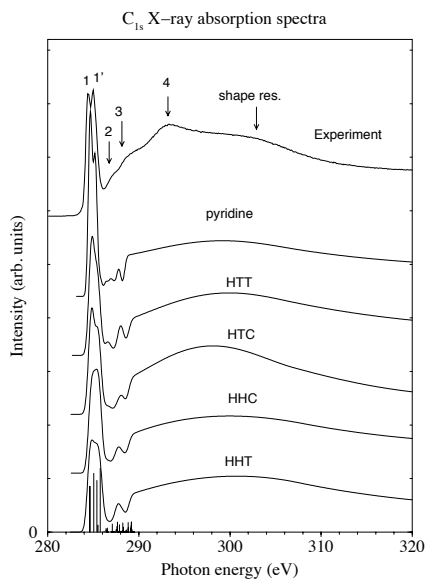

**Fig. 3:** Experimental and calculated XA spectra of PPy at the $C_{1s}$ threshold, where the calculated spectra have been aligned to the $\pi^*$ peak of the experimental spectrum.

## 4.2 Resonant and nonresonant X-ray emission spectra

Figure 4,5,6,7 show non-resonant and resonant X-ray emission spectra of poly-pyridine. The energy scales of the spectra have been aligned by using the elastic peak in the resonant spectrum excited at the $\pi^*$ peak in the absorption spectrum. The corresponding calculated non-resonant and resonant X-ray emission spectra respectively, are also presented in these figures. Five peaks (labeled A-E) can be identified in these spectra. Peak A corresponds to $\pi$-electron and $n$ electron states ($1a_2$, respectively, $11a_1(n)$ in pyridine), peak B is due to both $\pi$ and $\sigma$ with $\sigma$ dominating, while peaks C,D,E, all are due to $\sigma$ electronic states. The calculated spectra are shown for the four different isomeric types of poly-pyridine. Calculated spectra of the pyridine molecule are also shown for comparison. The five bands show up clearly in the calculation of the pyridine molecule as well as for the four different isomers of poly-pyridine which were obtained by taking the repeat-unit cell as a model molecule. In the case of poly-pyridine, the bands are broader due to a larger number of transitions involved in each band which is also the case for a real polymer system.

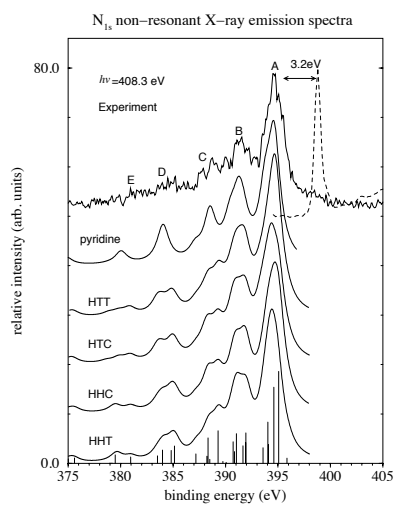

**Fig. 4:** Experimental and calculated non-resonant $N_{1s}$ XE spectra of PPy.

It is noteworthy that the bands A-E observed in the four types of X-ray emission spectra reported here, correspond well in energy with the five main bands (B-F) observed in the ultraviolet photoelectron spectra (UPS) of PPy reported by Miyamae et al. [3]. This is quite natural from the point of view that the same final states are obtained in the UPS and nonresonant C1s and N1s X-ray emission spectra. The intensities are evidently different owing to the different nature of the transition moments between the spectroscopies.

### 4.2.1 X-ray emission spectra at the $N_{1s}$ threshold

Figure 4 and 5 show N1s non-resonant and resonant X-ray emission spectra of poly-pyridine, excited with 408.3 and 398.8 eV photon energies, respectively. In the resonant calculation (Fig. 5), a similar band structure is seen as in the non-resonant case, but the resonant spectra show a stronger isomeric dependence. The latter fact can be referred to the interference effect which is operating for resonant emission, and which is very dependent on the actual electronic structure. In Fig. 5, the relative intensity of the B-band is about 73% as high as that of *A*-band, for the HHC isomeric form, but about 45% for the other three isomers. Differences in intensity distribution of the C-bands are also predicted by the calculations. The comparison with the experimental spectrum seems to disfavour the HHC form and favour the HTT and HHT forms.

Chemical Physics **237**, 295 (1998)

### 4.2.2 X-ray emission spectra at the $C_{1s}$ threshold

Figure 6 and 7 show the experimental resonant and non-resonant X-ray emission spectra of poly-pyridine excited at 284.4 eV and 305 eV photon energies, respectively. The energy scale of the experimental spectra have been aligned together by using the strong C1s elastic peak in the resonant spectrum excited at the first $\pi^*$ peak in the absorption spectrum. The carbon spectra obviously map the same final levels as the nitrogen spectra but with different transition moments owing to the different intermediate states. Thus, while the carbon spectra show a similar peak structure of the bands, the intensity distribution is different. In both cases the resonant spectra show strong elastic (recombination) peaks.

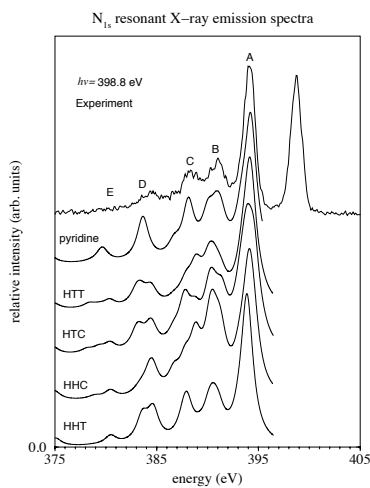

**Fig. 5:** Experimental and calculated resonant $N_{1s}$ XE spectra of PPy.

The non-resonant X-ray emission spectra are dominated by the C K-edge $2p \rightarrow 1s$ diagram transitions between the core and valence vacancy states. As seen in the non-resonant spectrum a peak is clearly visible at about 284 eV due to satellite transitions. These shake-up satellite transitions originate from multi-vacancy states and are traditionally referred to as Wentzel-Druyvesteyn (WD) satellites [31, 32].

The calculated $C_{1s}$ spectrum of the pyridine molecule shows a similar band structure compared to the $N_{1s}$ spectrum, but with a different intensity distribution among the bands, reflecting the localization of the core-hole orbitals. The most remarkable difference is the intensity of the A band. It becomes less intensive in $C_{1s}$ spectrum, because the n-lone pair MO's have little dipole overlap with localized $C_{1s}$ core orbital. On the other hand, the MO's of the B-band have larger 2p contributions from the carbon atoms and so a larger dipole overlap with the $C_{1s}$ core orbitals than with the $N_{1s}$ core orbitals. For the inner MO's, such as the E-band, the intensities are weak in both the $C_{1s}$ and $N_{1s}$ spectra due to the larger 2s character of the MO's and probably also due to the breakdown of the MO picture with accompanying correlation state splittings [33].

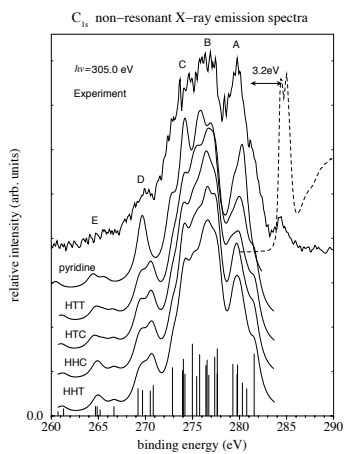

**Fig. 6:** Experimental and calculated non-resonant $C_{1s}$-edge XE spectra of PPy.

The simulated C1s non-resonant spectra for the different isomers show no essential difference between each other. However, the calculated resonant spectra for the different poly-pyridine isomers in Fig. 7 show an isomeric dependence. As in the nitrogen case the best experimental comparison is obtained for the HHT and HTT isomeric forms, c.f. Figs. 5 and 7. It can be noted that a comparison between the computed density-of-states distribution with the measured UPS spectrum of PPy by Miyamae et al., gives some preference to the HH form [3].

# 5 Discussion

Comparing the resonant and non-resonant spectra, the largest difference occurs at about 280 eV photon energy, where the *A*-band appears only as a weak feature in the resonant spectrum. A depletion of the "*A*-band" has previously been observed for resonant X-ray emission spectra of benzene [19] and is the result of the strict parity selection rule applied to the $D_{6h}$ symmetry of that molecule. In previous studies of aniline [20] and poly-phenylene-vinylene (PPV) polymers [15], it was argued that the multi-channel interference effects make transitions from π MO's of the *A*-band to the $C_{1s}$ core orbitals effectively forbidden. Thus the interference effect - and so the symmetry selectivity - grows progressively stronger as the chemical disturbance of the benzene rings becomes weaker. The *A*-band emerged thus only as a weak feature in both the experimental and calculated spectra of those hydrocarbon polymers.

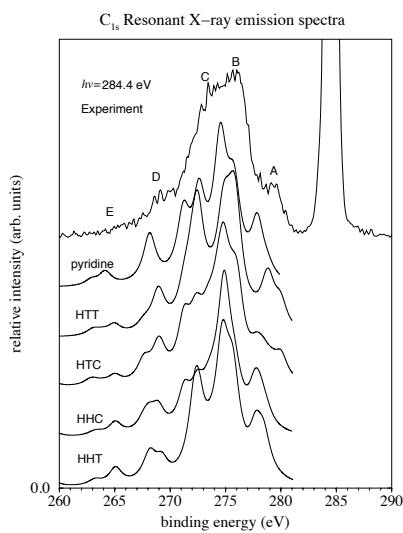

**Fig. 7:** Experimental and calculated resonant $C_{1s}$ XE spectra of PPy.

In the work on PPV [15] the connection between symmetry selection for RIXS observed in molecules and the momentum conservation rule of solids [34,35,36,37] was pointed out. Just as the symmetry selection in the resonant aniline spectrum the momentum conservation for resonant emission in the polymers is an effect of channel interference. For π-electron polymers the momentum conservation leads to depletion of emission from π-levels, as confirmed in the present work and in ref. [15] for the PPV compounds. In both cases a similar depletion could be obtained by full symmetry and interference analysis of the resonant process using the repeat unit as a model. One finds the depletion to be about as strong for poly-pyridine as for PPV compounds although one would expect a smaller effect for poly-pyridine due to the stronger chemical shifts of the core-excited states. We do not observe a corresponding depletion going from the non-resonant to the resonant spectra in the nitrogen case, which owes to the simple fact that the strong high-energy band in those spectra is dominated by the lone-pair *n* orbital localized to nitrogen, which has σ symmetry. So the different localization characters of the emitting levels are also very relevant to consider for the analysis of the resonant spectra.

In the calculated spectra a difference between the edge structure at the frontier (at the lower binding energy side of nonresonant C1s spectrum) of the *A*-band can be observed for the polymer compared to the pyridine molecule. This edge structure results from the transition of the highest occupied molecular orbital. By subtracting the energy of this edge structure in the non-resonant spectrum with that of the elastic peak of the resonant spectrum an alternative way of obtaining the band gap experimentally is obtained which also was demonstrated in our previous paper [15]. The band gap so obtained is 3.2 eV which agrees fairly well with the value of 2.82 eV estimated from the edge of the absorption spectra of solid fillm PPY [3].

# 6 Summary

Resonant and non-resonant X-ray scattering spectra of poly-pyridine have been measured with monochromatic synchrotron radiation. The spectra show structure rather far into the valence energy region. The emission bands are quite similar in the carbon and nitrogen spectra -the same final states are involved- but have different relative intensities in the two spectra, because the localization properties of the energy bands are different. In particular, the lone-pair *n* levels emphasize the high energy part of the nitrogen spectrum in comparison with the carbon spectrum. These qualitative features of the XE spectroscopy, well-known for free molecules, are here demonstrated for a polymer. The most salient difference between the resonant and non-resonant spectra is the partial depletion of the $\pi$ band in the carbon case in line with what can be expected from frozen orbital theory of resonant XE spectra of $\pi$-electron systems.

An analysis based on *ab initio* canonical Hartree-Fock theory reproduces the main features of the poly-pyridine spectra both for carbon and nitrogen core excitations. The simulations indicate some isomeric dependence of the resonant X-ray emission spectra, giving preference to two of the four isomers studied, while the gross profiles of the non-resonant X-ray emission and absorption spectra are quite independent of the isomeric forms; the $\pi^*$ transitions show some internal structure due to isomeric dependent chemical shifts.

# 7 Acknowledgments

This work was supported by the Swedish Natural Science Research Council (NFR), the Swedish Research Council for Engineering Sciences (TFR), the Göran Gustavsson Foundation for Research in Natural Sciences and Medicine and the Swedish Institute (SI). The experimental work at ALS, Lawrence Berkeley National Laboratory was supported by the director, Office of Energy Research, Office of Basic Energy Sciences, Materials Sciences Division of the U. S. Department of Energy, under contract No. DE-AC03-76SF00098. Computer time has been generously supported by NSC, the Swedish national supercomputer centre in Linköping. L.Y. would like to acknowledge the support from National Natural Science Foundation of China.